\documentclass{article}

\usepackage[preprint]{neurips_2025}

\usepackage[utf8]{inputenc} 
\usepackage[T1]{fontenc}    
\usepackage{hyperref}       
\usepackage{url}            
\usepackage{booktabs}       
\usepackage{amsfonts}       
\usepackage{nicefrac}       
\usepackage{microtype}      
\usepackage{xcolor}         

\usepackage{amsmath}
\usepackage{amssymb}
\usepackage{graphicx}
\usepackage{hyperref}
\usepackage{xcolor}
\usepackage{booktabs}
\usepackage{geometry}
\usepackage{listings}
\usepackage{enumitem}
\usepackage{csquotes}
\usepackage{float}
\usepackage{algorithm}
\usepackage{algorithmic}
\usepackage{xspace}

\title{Seeing the Forest Through the Trees: Knowledge Retrieval for Streamlining Particle Physics Analysis}
\workshoptitle{Machine Learning and the Physical Sciences}

\author{%
  James McGreivy,
  Blaise Delaney,
  Anja Beck,
  Mike Williams \\
NSF AI Institute for Artificial Intelligence and Fundamental Interactions \\
MIT
}

\newcommand{\GraphRAG}{\textsc{SciGraphRAG}\xspace}
\newcommand{\TreeRAG}{\textsc{SciTreeRAG}\xspace}

\begin{document}

\maketitle

\begin{abstract}
Generative Large Language Models (LLMs) are a promising approach to structuring knowledge contained within
the corpora of research literature produced by large-scale and long-running scientific collaborations.
Within experimental particle physics, such structured knowledge bases could expedite methodological and editorial review. Complementarily, within the broader scientific community, generative LLM systems grounded in published work could make for reliable companions allowing non-experts to analyze open-access data.
Techniques such as Retrieval Augmented Generation (RAG) rely on semantically matching localized text chunks, but struggle to maintain coherent context when relevant information spans multiple segments, leading to a fragmented representation devoid of global cross-document information.
Here, we utilize the hierarchical organization of experimental physics articles to build a tree representation of the corpus, and present the \TreeRAG system that uses this structure to create contexts that are more focused and contextually rich than standard RAG.
Additionally, we develop methods for using LLMs to transform the unstructured corpus into a structured knowledge graph representation. We then implement \GraphRAG, a retrieval system that leverages this knowledge graph to access \emph{global} cross-document relationships eluding standard RAG, thereby encapsulating domain-specific connections and expertise.
We demonstrate proof-of-concept implementations using the corpus of the LHCb experiment at CERN.
\end{abstract}

\section{Introduction}

LLMs and the systems built around them are rapidly transforming the way that technical knowledge is organized and retrieved.
The format of experimental physics articles, however, remains largely unchanged, still consisting of text, tables, and visual representations of data such as histograms packaged into a rigid structure and format.
In this context, large
research collaborations, \emph{e.g.}\  those at the Large Hadron Collider (LHC),
have produced thousands of publications, along with various types of supplemental data files containing the values of measurements and other types of results.

Crucially, many CERN-based collaborations have made subsets of their expert-curated
data available for public use~\cite{CERNOpenDataPolicy, CERNOpenDataPortal2024}.
While the knowledge required to analyze these public data is, in principle, contained in the
experiment's corpus of articles, in practice reading such a large corpus is prohibitively time consuming.
More critically, synthesizing the requisite analysis techniques requires tracing interconnected knowledge dependencies distributed across hundreds of articles, effectively requiring non-specialists to develop domain expertise.
Consequently, CERN open data remains largely underexplored despite its scientific potential.
In the same spirit, new members joining a large collaboration---such as incoming PhD students---face a steep learning curve
before they can effectively contribute.
Lastly, the sheer size of such collaborations fragments knowledge into sub-domain clusters, often leading to duplicated efforts and delayed progress due to limited cross-group awareness.

These bottlenecks suggest that automated knowledge-retrieval systems could significantly lower entry barriers.
Indeed, studying LLM-based approaches for scientific knowledge retrieval is an active area~\cite{beltagy-etal-2019-scibert,han2021pubmedbert,he2025biorag,Xu:2025uwt,buehler2024materialrag,zhong2025benchmarking}, including at the LHC where the focus has been on broader organizational tools~\cite{chatlas,fm4s}.
The tendency of LLMs to {\em hallucinate} is generally mitigated using
RAG techniques~\cite{rag_in_llms},
which involves retrieving relevant text chunks from a pre-chunked corpus in order to populate the context of the LLM.
There are a few problems, however, with using traditional RAG for this task:

\textbf{Accidental Semantic Similarity} Traditional RAG systems rely solely on the reductive metric of chunk-based semantic similarity, which can lead to retrieval of irrelevant passages that happen to share keywords or phrasing with the query but lack topical relevance or broad contextual awareness.

\textbf{Fragmented Context}
Traditional RAGs fragment context by concatenating text chunks based purely on semantic similarity, without regard for their logical relationships or the original document structure.
This approach can juxtapose unrelated passages, creating incoherent information dumps that mislead LLMs into assuming false connections between adjacent but unrelated content.

\textbf{Lack of Global Knowledge}
Expert-level responses often require \textit{global} knowledge, emergent from the relationships and patterns that become apparent only when the corpus is considered as a whole.
Traditional RAG systems are limited to retrieving isolated chunks of text based on semantic similarity, preventing access to this emergent layer of global knowledge.

Accidental semantic similarity is typically addressed with rerankers~\cite{nogueira2019passage, nogueira2020document}, which use cross-encoder architectures to jointly encode query–document pairs and reorder text chunks based on richer semantic signals than embedding similarity.
However, this \textit{post-hoc} fix  overlooks knowledge fragmentation, as it evaluates chunks in isolation without regard for their role in the document’s logical structure.

In this work, we address these limitations through two complementary approaches built from the corpus of the LHCb collaboration.
We introduce \TreeRAG, a knowledge-retrieval method that leverages the hierarchical and broadly uniform structure of LHCb articles to enable focused and contextually rich retrieval while avoiding fragmenting information.
Furthermore, we
develop
a knowledge graph of LHCb-specific systematic uncertainties and analysis methods through LLM-inferred, context-aware connections. This \GraphRAG approach encapsulates the layer of global knowledge inaccessible to traditional RAG, acting as a queryable surrogate of domain expertise.

\section{\TreeRAG: Local Knowledge Retrieval}
\label{sec:TreeRAG}

\TreeRAG addresses the issues of accidental semantic similarity and fragmented context by exploiting the hierarchical tree-like structure inherent to scientific publications.
For each article, it builds a tree representation to guide retrieval toward semantically relevant content in contextually appropriate sections.
The tree representation also lends itself to the creation of more structured contexts helping prevent the issue of context fragmentation.

The tree-construction process works as follows: the (sanitized) \LaTeX\ source is parsed to find hierarchical elements (sections, subsections) and indivisible units (paragraphs, figure captions, table captions, equations).
Each article is represented by a tree with the abstract at the root node, each section and subsection creating intermediate nodes, and the atomic content at the leaf nodes.
All intermediate nodes are given brief LLM-generated summaries by recursively concatenating and summarizing the summaries of their children.
The root node summary is the abstract, and the base case for a leaf node summary is its atomic content.
While this recursive summarization incurs a moderate upfront computational cost, it is performed only once during tree creation, making subsequent retrieval operations efficient.
See Appendix \ref{app:TreeRAG} for details on how the tree representation is constructed, how relevant information is retrieved from it, and for computational costs.

Each node is initially assigned a dense vector embedding by feeding its summary into a paragraph embedding model.
These embeddings are refined through a recursive attention-weighted process that leverages the hierarchical structure to filter spurious matches.
Since section summaries and their constituent content represent the same information at different abstraction levels, the hope is to amplify semantic signals robust across both representations while diluting incidental features like word choice artifacts or summary hallucinations that might cause false retrieval matches.

\TreeRAG retrieves information by traversing the document hierarchy from abstract to specific content.
Rather than retrieving chunks based solely on semantic similarity, \TreeRAG selects chunks that are both semantically similar to the query---and originate from sections that are topically relevant to the query.
The retrieval algorithm uses a greedy tree traversal that prioritizes the most promising document sections before examining their constituent chunks.
The refined embeddings help avoid exploration in topically irrelevant directions incidentally semantically similar to the query.

\section{\GraphRAG: Global Knowledge Retrieval}

Although \TreeRAG excels at retrieving focused information within individual documents, it suffers from the standard RAG inability to access relational information or patterns emergent from the entire corpus.
To address this, \GraphRAG uses an LLM-constructed knowledge graph (KG) representation of the corpus to store structured relational information. The system uses \textsc{Cypher}, the Neo4j graph querying language~\cite{neo4j}, to make this KG representation queryable.

KGs require a \textit{schema}, which defines the graph structure and the types of entities, relationships, and attributes it can contain.
The LHCb schema is based on the structure of particle physics analyses---in particular the determination of measurement uncertainties.
This schema focuses on how different analyses handle similar systematic effects; which uncertainties dominate for specific types of measurements; and how methods for treating uncertainties have evolved over time.
In addition, the schema is concise and narrowly scoped to enable being described to an LLM.

Effective KG construction requires identifying and linking entities and relationships across the full corpus, which is challenging due to LLMs being constrained by finite context windows.
To address this, we leverage the fact that high-quality KGs can be constructed from individual articles, whose content typically fits within the LLM context window.
We first generate per-article KGs then subsequently perform cross-document canonicalization~\cite{Zhang2024ExtractDC} to produce a cohesive, corpus-level KG with high interconnectivity.

To construct per-article KGs, an LLM is first fed the abstract along with instructions to extract high-level information about the analysis, measured observables, and relevant physical processes.
Then, the LLM is fed relevant text from the body of the article to extract a KG representation of the key sources of uncertainty and the methods used to  determine them.
A modern LLM such as GPT-5 mini is capable of undertaking these tasks, and in our observation can produce accurate graph representations of an analysis.
Similarly to \TreeRAG, this step does incur a moderate upfront computational cost only once during its creation.
However, the transformation of the corpus into this structure allows access to information which would otherwise be extremely difficult to query.

The next step, \emph{canonicalization}, assimilates the individual KGs into a single unified KG.
This process resolves duplicate entities across individual KGs by iteratively combining similarity clustering with LLM-as-judge merge decisions.
Since each KG may have thousands of entities, similarity clustering is used as an initial filter to reduce the complexity of the canonicalization.
All KG entities are transformed into a hybrid vector representation, with TF-IDF used to vectorize the entity names (where keywords dominate the meaning) and semantic embeddings used for entity descriptions (where semantic content dominates), which are then combined into a weighted similarity score.
Entities of the same type are agglomeratively clustered in this combined similarity space, with constraints applied to prevent merging entities from the same source paper.
The clusters are then fed asynchronously to LLM judges which make the final decision on which entities within a cluster are duplicates and perform the necessary merging.
(See Appendix \ref{app:GraphRAG} for computational costs.)

As we show in Sec.~\ref{sec:Experiment}, this canonicalized KG constructed from the entire LHCb corpus stores \textit{global knowledge} that was previously inaccessible by querying on unstructured text.
To access this knowledge, natural language (NL) queries must be translated into the \textsc{Cypher} graph-querying language, for which we currently use an LLM that
is knows the graph schema, as well as policies to follow when constructing an effective query.
The resulting \textsc{Cypher} expression defines cuts on the graph, which are evaluated in order to return a subgraph in response to the query.
The returned subgraph undergoes LLM-based synthesis to produce an NL answer to the original NL query.

\section{Experiments}
\label{sec:Experiment}

We compare \TreeRAG to standard RAG via a systematic study using the HFLAV evaluation dataset described in Sec.~\ref{ch:evaluation:HFLAV}.
The evaluation methodology employs an LLM-as-judge approach with instructions to answer the query using only RAG-provided context.
Figure~\ref{fig:quality_proportions} shows the distribution of response quality grades. \TreeRAG demonstrates modest but consistent improvements over RAG, receiving "poor" ratings only 10\% of the time compared to  25\% for RAG. In addition, \TreeRAG achieves "satisfactory" or "good" ratings more than 50\% of the time, compared to approximately 42\% for standard RAG.
(Full details provided in App.~\ref{app:base-vs-tree}.)

The \GraphRAG system needs more development to make automated KG construction and NL-to-Cypher translation reliable.
Thus, we only demonstrate its performance on human-vetted example queries from the HFLAV dataset in App.~\ref{app:graph-ex}, with one example provided here.

\textbf{Query $\gamma$} What decay channel should I use to measure the CP violating phase \texorpdfstring{$\mathbf{\gamma}$}{gamma}? What sources of systematic uncertainty will dominate my analysis?

\textbf{\GraphRAG Response}
\textit{
Use a combination of tree-level $B \to D^{(*)} h$ modes plus the time-dependent $B_s \to D_s^{\mp}K^{\pm}$ channel. Practically this means including $B^{\pm} \to D K^{\pm}$ and $B^{\pm} \to D \pi^{\pm}$ (with $D \to K_S \pi \pi$, $K_S K K$, $K^{\pm} \pi^{\mp}$, $K^{\pm} \pi^{\mp} \pi^{\pm} \pi^{\mp}$, 3- and 4-body final states), $B^{\pm} \to D^* K^{\pm}$, $B^0 \to D K^{*0/\pm}$ and $B^0 \to D K^+ \pi^-$ (GLW/ADS/GGSZ-type approaches), and the time-dependent $B_s^0 \to D_s^{\mp} K^{\pm}$ analyses --- combining these channels is the route taken in recent LHCb determinations and gives the best precision on $\gamma$ [2010.08483, 2410.21115, 2209.03692, 2412.14074].}

\textit{The dominant systematic uncertainties will be external hadronic inputs and model/theory assumptions: uncertainties on $D$-decay hadronic parameters ($r_D^X$, $\delta_D^X$, $\kappa_D^X$, CP-even fractions $F_+$ for multi-body modes) and the coherence/strong-phase inputs used in model-independent Dalitz/binning approaches (these appear as a separate ``external/strong-phase'' uncertainty in several measurements) will dominate; if you use U-spin relations the non-factorizable U-spin-breaking modelling is a large, non-reducible theory systematic; and using external constraints (e.g.\ world-average $\gamma$ or $-2\beta_s$ as inputs) propagates their uncertainties into your result [2010.08483, 1408.4368, 2311.10434].}

\begin{figure}
    \centering
    \includegraphics[width=0.4\linewidth]{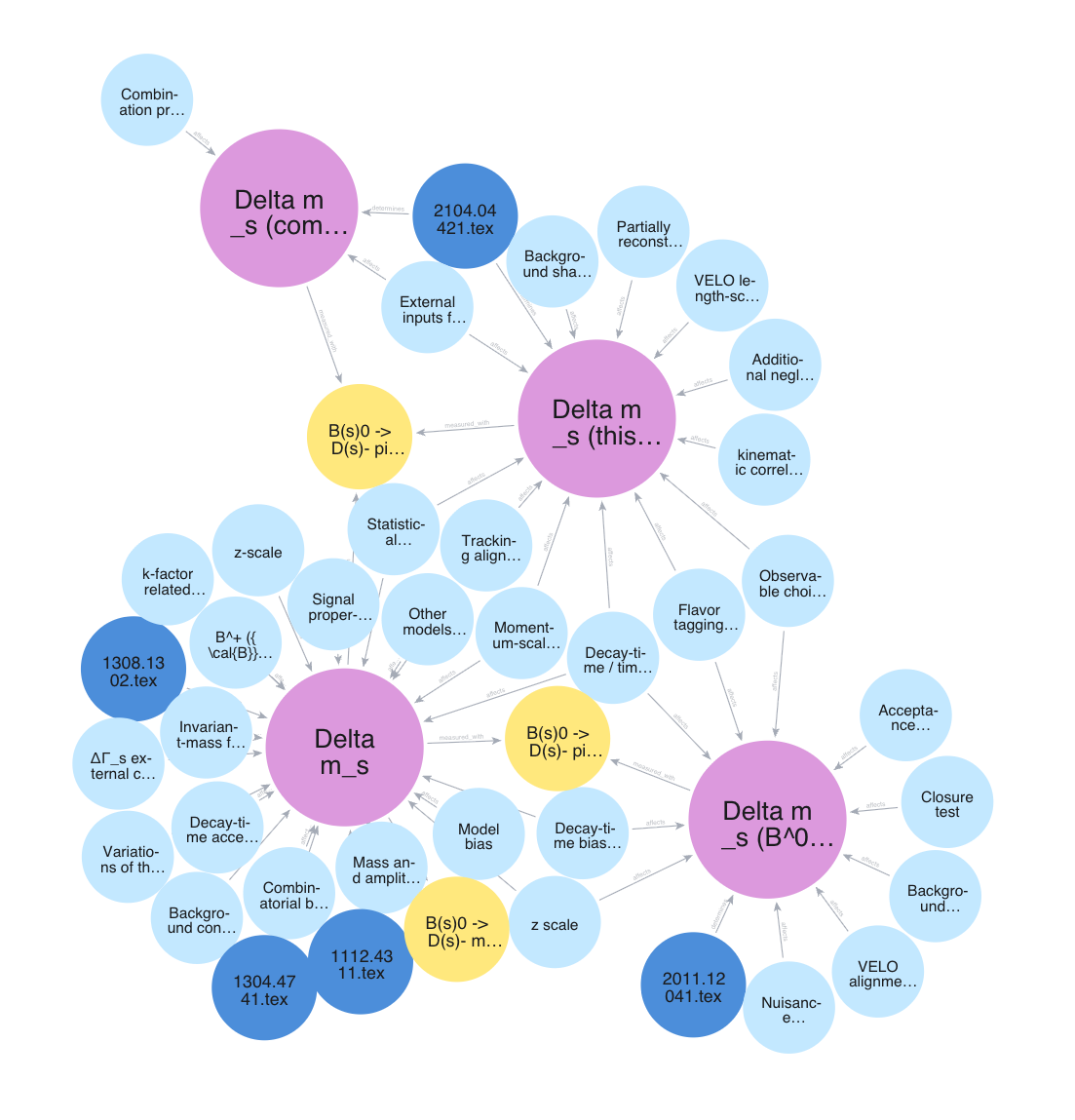}
    \includegraphics[width=0.44\linewidth]{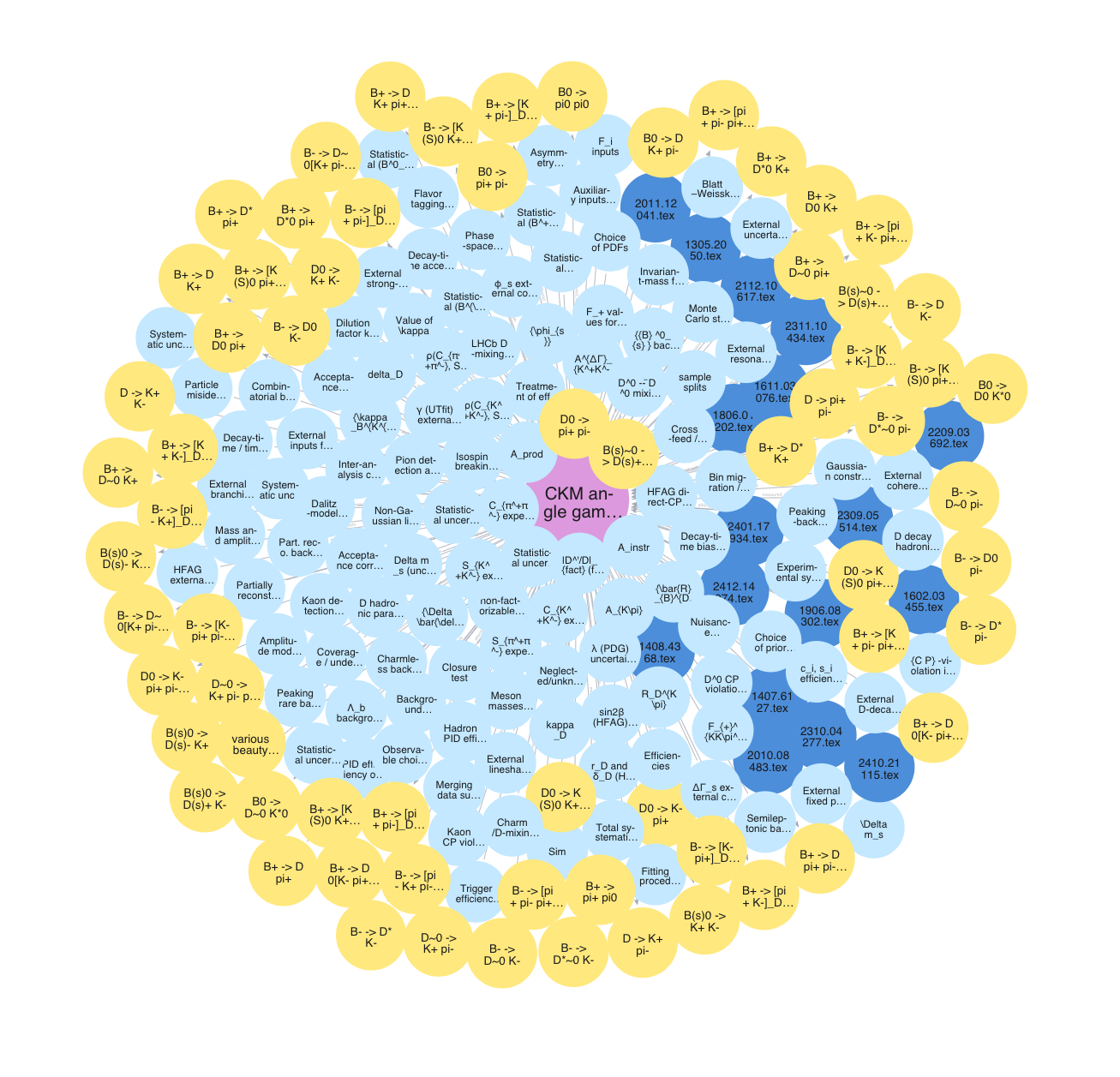}
    \caption{Visualization of knowledge subgraphs for Query (left) $\Delta m_s$ and (right) $\gamma$ (see  App.~\ref{app:graph-ex})
    showing relationships between articles (dark blue), decay processes (yellow), observables (purple), and uncertainty sources (light blue). The four duplicated "Delta m\_s" entities in the left graph are evidence of imperfect entity resolution during KG construction. In the right graph, all 17 "CKM Angle gamma" observables from distinct articles were properly canonicalized into a single entity.}
    \label{fig:query_graph}
\end{figure}

The LLM-as-judge score for this response is "good" whereas both the \TreeRAG and RAG are rated only "satisfactory" for this query.
(We agree with these ratings.)
Examples like this serve as illustrative cases of what a more mature \GraphRAG implementation could achieve.
Figure~\ref{fig:query_graph} shows that the KG for this query contains all LHCb $\gamma$ measurements, along with the relevant decay processes and uncertainty sources.
(Note that the code will be made public and linked in the camera-ready version of this workshop paper.)

\section{Broader Impact}
Automatic knowledge retrieval via LLMs is a topic with clear broader impacts across the field of experimental physics.
As discussed above, we believe that knowledge graphs and graph clustering provide a path to producing expert-level responses by making it possible for the LLM to \textit{see the forest through the trees}.
A key component of this approach is producing knowledge graphs for each article.
We believe that the best approach here is to first agree on a common high-level schema to be used across all LHC experiments.
This will not only make it easier to develop common tools for automatic knowledge retrieval, it will also make these knowledge graphs human understandable across the experiments.
For future articles, ideally knowledge graphs would be produced and vetted by the collaborations and published with the articles.
For past articles, which number in the thousands, these can be auto produced (as we have done here) but at a minimum they should be validated by the collaborations.
For the LHC experiments, this approach could enable producing an analysis co-pilot to help non-experts analyze their open data or to help their own PhD students analyze the full (private) data samples.
We also note that past experiments, such as those from LEP, could also benefit from such a tool making analyzing their open data much easier.

\begin{ack}
This work was supported by NSF grant PHY-2019786 (The NSF AI Institute for Artificial Intelligence and Fundamental Interactions, http://iaifi.org/).
\end{ack}

\bibliographystyle{unsrtnat}
\bibliography{main}

\appendix

\section{Evaluation Datasets}

\subsection{LHCb Corpus}
\label{ch:evaluation:LHCbCorpus}

For this work a corpus of 834 LHCb publications~\cite{cern_alcm_analysis} was assembled by querying the \textsc{Inspire-HEP}~\cite{inspirehep} literature database API. This collection comprises published and peer-reviewed measurement papers, detector-performance papers, conference contributions, review papers, and theoretical papers published by the LHCb collaboration between December 2009 and August 2025.

For each publication, the raw \LaTeX~source was downloaded and merged into a monolithic source file using the \texttt{latexpand} command-line toolkit~\cite{ctan:pkg:latexpand}. This document was further processed to remove all extraneous \LaTeX~content, comments, bibliography entries, and collaboration author lists. All LHCb collaboration and user-defined latex macros were expanded down to raw \LaTeX~source using the \texttt{expand-latex-macros} Python library~\cite{pypi:expand-latex-macros}.

\subsection{HFLAV Eval Q\&A}
\label{ch:evaluation:HFLAV}

In order to evaluate the effectiveness of our RAG systems, we need a ground truth dataset of questions with known answers that could reasonably be asked from the LHCb corpus. To this end, we developed an evaluation strategy leveraging the  Heavy Flavour Averaging Group (HFLAV) report "Averages of $b$-hadron, $c$-hadron, and $\tau$-lepton properties as of 2023" \cite{HFLAV}, a meta-analysis devoted to the statistical combination of heavy-flavour results including many from the LHCb collaboration.

The HFLAV report was specifically chosen in light of following considerations: (1) it is not included in the LHCb corpus, ensuring our evaluation tests genuine retrieval capabilities rather than memorization; (2) it contains detailed technical content requiring cross-document synthesis from multiple LHCb analyses; (3) it provides verifiable numerical benchmarks and methodological comparisons that can serve as objective accuracy metrics.

For dataset construction we separately processed each HFLAV chapter containing LHCb-relevant content. Tables, measurement summaries, and technical discussions were parsed by GPT-5 mini to identify key physics concepts, measurement techniques, and numerical results that would be expected to appear in a comprehensive LHCb corpus analysis.

Using GPT-5 mini, we generated evaluation queries designed to test both precision retrieval (finding specific technical details) and synthesis capabilities (understanding relationships across multiple analyses). The LLM was prompted to specifically target queries about optimal decay channels, systematic uncertainty sources and evaluation protocols, and methodological comparisons across analyses. Each query also needs to include factual checkpoints (specific decay channels, numerical uncertainties, measurement techniques) that can be objectively evaluated against HFLAV reference values

Queries are generated alongside a detailed rubric:
\begin{itemize}
    \item \textbf{Essential requirements:} Core physics principles and basic methodology that any competent answer must include;
    \item \textbf{Expert-level requirements:} state-of-the-art methodologies for systematic uncertainty evaluation, cross-analysis correlations, and validation methods that distinguish comprehensive responses;
    \item \textbf{Factual benchmarks:} Specific numerical values, decay processes, and procedural details that can be verified against the HFLAV reference.
\end{itemize}

This process generated 8 queries per chapter across 7 relevant HFLAV chapters, resulting in a 56-question evaluation dataset covering a breadth of LHCb physics topics. Example queries include:

\begin{quote}
    \textbf{Query:} What are the most precise measurements of $b$-hadron lifetimes performed by LHCb, and how do the systematic uncertainties in these measurements compare across different decay channels and methodologies?
\end{quote}

\begin{quote}
    \textbf{Query:} What is LHCb's most precise measurement of $|V_{cb}|$ from exclusive semileptonic $B$ decays, and how does it compare to the inclusive determination? What are the dominant sources of theoretical and experimental uncertainty in the exclusive approach?
\end{quote}

\begin{quote}
    \textbf{Query:} Which $B^0$ decay mode has been most useful for studying polarization fractions in vector-vector final states, and what are the LHCb results?
\end{quote}

\section{Additional Experiments}

\subsection{Standard RAG vs \TreeRAG}
\label{app:base-vs-tree}

To evaluate the effectiveness of \TreeRAG against standard RAG approaches, we conducted a systematic comparison using the HFLAV evaluation dataset described in Section \ref{ch:evaluation:HFLAV}. We tested three configurations: \textsc{BaseRAG} (standard chunking-based retrieval), \TreeRAG without diffusion, and \TreeRAG with diffusion-enhanced embeddings. The analysis was averaged across multiple context window sizes (8k, 16k, and 32k tokens), with no significant dependence found on context window size.

The evaluation methodology employed an LLM-as-judge approach where the context from all three RAG systems was inputted to an LLM, with instructions to answer the query using only the provided context. These answers were anonymized and shuffled before being simultaneously presented to an evaluator LLM along with the detailed rubrics developed for each query. The evaluator assigned grades on a four-point scale: "Poor", "Below Average", "Satisfactory", and "Good", based on objective metrics within the query rubric.

A human expert validation was performed on a subset of the LLM-generated grades, finding no evidence of significant hallucinations and good adherence to the established rubrics. However, since this human-expert validation was not conducted across the entire evaluation dataset, these results should be interpreted with appropriate caution.

\begin{figure}
    \centering
    \includegraphics[width=0.8\linewidth]{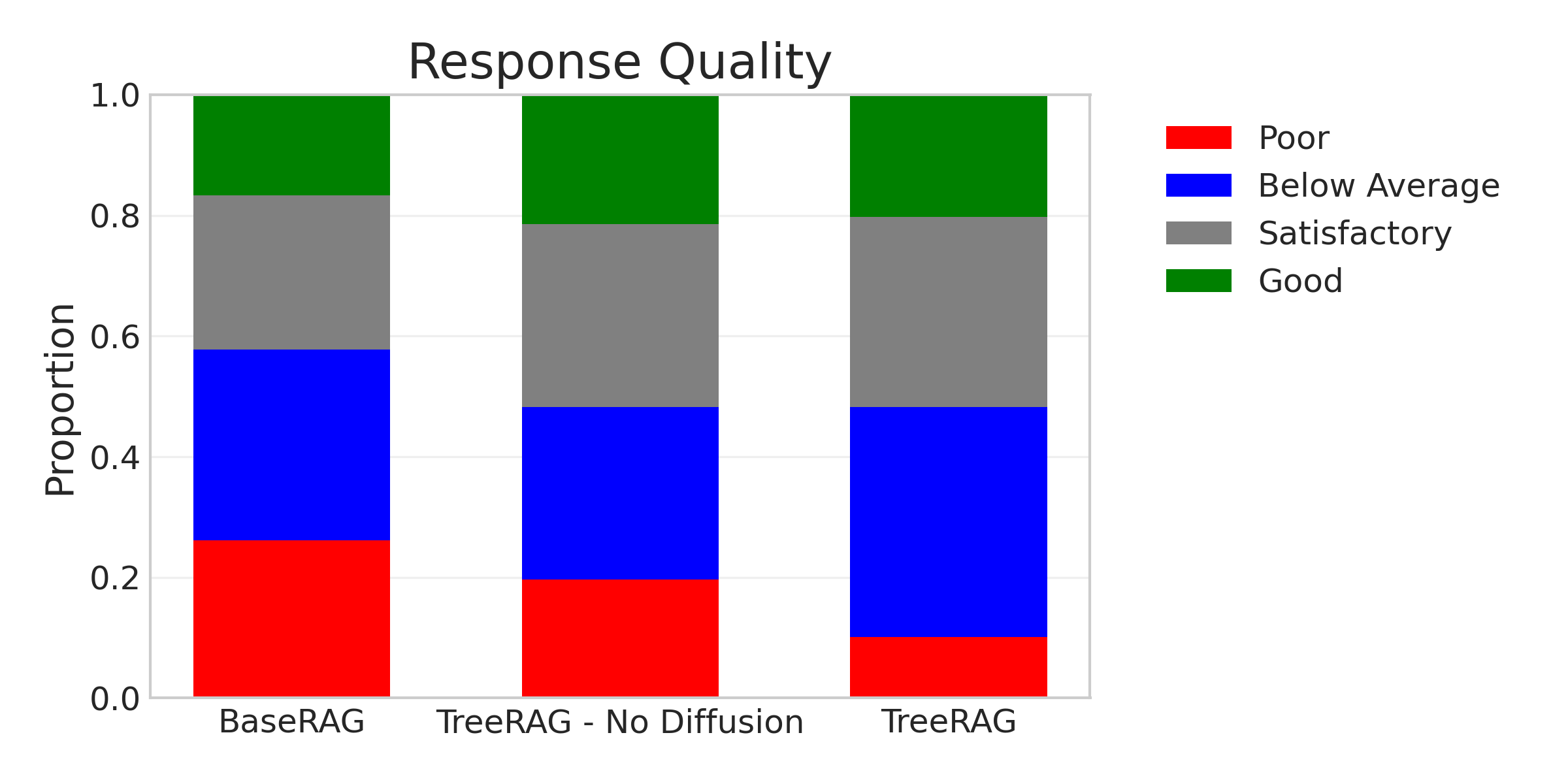}
    \caption{Distribution of response quality grades across \textsc{BaseRAG}, \TreeRAG without diffusion, and \TreeRAG with diffusion systems evaluated on the HFLAV dataset. Results are aggregated across context windows of 8k, 16k, and 32k tokens.}
    \label{fig:quality_proportions}
\end{figure}

Figure \ref{fig:quality_proportions} presents the distribution of response quality grades across the three systems. \TreeRAG demonstrates modest but consistent improvements over BaseRAG. Most notably, \TreeRAG with diffusion achieves the best performance, receiving "poor" ratings only 10\% of the time compared to 20\% for \TreeRAG without diffusion and 25\% for BaseRAG. Both \TreeRAG configurations achieve "satisfactory" or "good" ratings more than 50\% of the time, compared to approximately 42\% for BaseRAG.

\subsection{\GraphRAG Examples}
\label{app:graph-ex}

The \GraphRAG system is still a work in progress, with room for improvement across the per-paper KG construction process, the graph canonicalization process, and the natural language to Cypher translation process. In addition, the current graph schema is narrowly scoped and only accommodates specific styles of query effectively. Our observation is that when GraphRAG is able to produce a sensible answer, it generally produces a very high quality one. However, GraphRAG also frequently fails due to poor mapping of the query onto the schema of the knowledge graph. Thus, we chose not to include the \GraphRAG in the systematic evaluation on the HFLAV dataset and instead demonstrate what it looks like when the \GraphRAG system is indeed successful.

Here we will demonstrate the performance of the \GraphRAG system on two queries, taken from the HFLAV eval dataset, which were able to map effectively onto the existing knowledge graph schema. The performance of the Base RAG and \TreeRAG systems were also included for completeness, but these results are not meant to serve as a direct comparison between the systems.

\begin{quote}

\textbf{Query $\Delta m_s$:} Which decay provides the most precise measurement of the \texorpdfstring{$\mathbf{B_s^0}$}{Bs} eigenstate oscillation frequency, \texorpdfstring{$\mathbf{\Delta m_s}$}{Delta}? What are the dominant systematic uncertainties for that measurement?

\textbf{Rubric:}
\begin{itemize}
    \item \textbf{Essential Requirements:}
    \begin{itemize}
        \item Recognize that $B_s^0 \to D_s^- \pi^+$ and $B_s^0 \to D_s^- \pi^+ \pi^- \pi^+$ are the golden channels for $\Delta m_s$ precision due to clean reconstruction and large branching fractions
        \item Understand that flavor tagging efficiency and mistag rates are critical systematic limitations
        \item Note that detector length scale calibration and momentum scale uncertainties dominate systematic errors
    \end{itemize}
    \item \textbf{Expert-Level Requirements:}
    \begin{itemize}
        \item Explain that the most precise measurements come from time-dependent oscillation analyses of flavor-tagged $B_s^0$ decays
        \item Recognize that non-flavor-specific modes like $B_s^0 \to J/\psi K^+ K^-$ can also measure $\Delta m_s$ but with larger uncertainties due to angular analysis complexity
    \end{itemize}
    \item \textbf{Factual Benchmarks:}
    \begin{itemize}
        \item The world average is $\Delta m_s = 17.766 \pm 0.004 \pm 0.004$ ps$^{-1}$
        \item LHCb's $B_s^0 \to D_s^- \pi^+$ measurement with 6 fb$^{-1}$ Run 2 data gives $17.768 \pm 0.005 \pm 0.003$ ps$^{-1}$
        \item LHCb's $B_s^0 \to D_s^- \pi^+ \pi^- \pi^+$ measurement with 9 fb$^{-1}$ combined data gives $17.757 \pm 0.007 \pm 0.008$ ps$^{-1}$
        \item Systematic uncertainties are dominated by: length scale ($\sim$0.003--0.006 ps$^{-1}$), momentum scale ($\sim$0.002--0.004 ps$^{-1}$), and fit bias ($\sim$0.002--0.008 ps$^{-1}$)
        \item CDF provided the first observation with hadronic and semileptonic modes yielding $17.77 \pm 0.10 \pm 0.07$ ps$^{-1}$
        \item The decay channels $B_s^0 \to D_s^- \pi^+$ and $B_s^0 \to D_s^- \pi^+ \pi^- \pi^+$ provide the most precise individual measurements due to clean flavor tagging and complete kinematic reconstruction
    \end{itemize}
\end{itemize}
\end{quote}

The following answers were constructed from the context provided by each of the three RAG systems:

\begin{quote}
\textbf{GraphRAG:}
The most precise direct measurements of the $B_s^0$--$\overline{B}_s^0$ oscillation frequency come from hadronic $B_s^0 \to D_s^- \pi^+$ (and closely related multi-hadron $D_s^- \pi^+ \pi^+ \pi^-$ control) channels. For example, the recent LHCb result quoted in the context gives $\Delta m_s = 17.7683 \pm 0.0051 \, (\text{stat}) \pm 0.0032 \, (\text{syst}) \, \text{ps}^{-1}$ from $B_s^0 \to D_s^- \pi^+$ [2104.04421]; earlier LHCb measurements on the same channels report compatible values with larger uncertainties (e.g.\ $17.768 \pm 0.023 \pm 0.006$ [1304.4741], and $17.757 \pm 0.007 \pm 0.008$ from a $D_s^- \pi^+ \pi^+ \pi^-$ control channel [2011.12041]). A combined LHCb ``legacy'' value is also quoted as $17.7656 \pm 0.0057 \, \text{ps}^{-1}$ (combination of these inputs) [2104.04421].

The dominant systematic effects identified in the provided context for these $\Delta m_s$ measurements are those that bias the reconstructed decay time: (i) decay-time biases from tracking/vertexing and PV--DV resolution (listed up to $\mathcal{O}(0.007)\,\text{ps}^{-1}$ in the context for time-bias studies) [2011.12041, 1308.1302], (ii) the detector $z$-scale / length-scale calibration along the beam axis (quoted $\simeq 0.004 \,\text{ps}^{-1}$ impact on $\Delta m_s$), and (iii) residual VELO alignment/misalignment effects ($\approx 0.0031 \,\text{ps}^{-1}$ in the context). For the combined legacy number, the combination procedure and assumed correlations between inputs (momentum/length scale, misalignment) are an additional external systematic/assumption that affects the quoted combined uncertainty but whose separate contribution is not specified in the provided context [2104.04421].

\textbf{TreeRAG:}
The most precise determinations of $\Delta m_s$ in the provided context come from LHCb measurements using $B_s^0 \to D_s^- \pi^+$ decays. The 2011 LHCb analysis (1.0 fb$^{-1}$, $\sim 34{,}000$ signal decays) reports $\Delta m_s = 17.768 \pm 0.023 \, (\text{stat}) \pm 0.006 \, (\text{syst}) \,\text{ps}^{-1}$ and is described as the most precise measurement to date [1304.4741]. (An earlier 2010 LHCb result using $B_s^0 \to D_s^- \pi^+$ and $B_s^0 \to D_s^- \pi^+ \pi^- \pi^+$ had a much larger statistical uncertainty: $17.63 \pm 0.11 \pm 0.02 \,\text{ps}^{-1}$ [1112.4311].)

The dominant systematic uncertainties identified in these LHCb analyses are small (the 2011 result has a $0.006 \,\text{ps}^{-1}$ systematic) and arise principally from the decay-time modelling: calibration of the per-event decay-time resolution (the $S_{\sigma_t}$ scale factor and its variation) and the decay-time acceptance function derived from full detector simulation. These sources are treated explicitly in the fits (resolution convolution with the scaled per-event $\sigma_t$ and an acceptance $\varepsilon(t)$ from simulation) and drive the quoted systematic uncertainty, which remains much smaller than the statistical error for the current datasets [1112.4311, 1304.4741].

\textbf{BaseRAG:}
The most precise measurement in the supplied context comes from LHCb using $B_s^0 \to D_s^- \pi^+$ decays (five $D_s^-$ channels): $\Delta m_s = 17.768 \pm 0.023 \, (\text{stat}) \pm 0.006 \, (\text{syst}) \,\text{ps}^{-1}$ [1304.4741]. An earlier LHCb result using $B_s^0 \to D_s \pi$ and $B_s^0 \to D_s 3\pi$ decays with a much smaller data set gave $\Delta m_s = 17.63 \pm 0.11 \, (\text{stat}) \pm 0.02 \, (\text{syst}) \,\text{ps}^{-1}$, but that measurement was completely statistically limited [1112.4311].

The dominant systematic uncertainties reported or discussed in these analyses are related to the decay-time resolution and its calibration (per-candidate estimated uncertainty $\to$ true $\sigma_t$ mapping) and small decay-time biases arising from selection and detector alignment. Several of the papers describe calibrating the per-candidate time uncertainty on prompt samples, modeling the resolution (often with scaled or double-Gaussian forms), and observing a small bias of order a few fs that must be included in the fit [1712.07428, 2011.12041, 1204.5675].
\end{quote}

For this query the \GraphRAG system finds observables that are physical constants with names containing "m\_s" and calculates their semantic similarity to the sentence "B\_s eigenstate oscillation frequency", keeping only those with similarity above 0.5. For these matching observables, it then traces through the experimental chain to find the scientific papers that determined these values, the particle decay processes used to measure them, and any other related observables involved in those measurements. It also identifies uncertainty sources that affect these measurements and calculates average uncertainty rankings. Finally, it returns different sources of experimental uncertainty which impact B\_s oscillation frequency measurements, along with the relevant papers and decay processes involved, all ordered by uncertainty ranking from most to least dominant.

The following answers were constructed from the context provided by each of the three RAG systems:

\begin{quote}
\textbf{Query $\gamma$:} What decay channel should I use to measure the CP violating phase \texorpdfstring{$\mathbf{\gamma}$}{gamma}? What sources of systematic uncertainty will dominate my analysis?

\textbf{Rubric:}
\begin{itemize}
    \item \textbf{Essential Requirements:}
    \begin{itemize}
        \item Definition of $\gamma$ as the CP-violating phase $\gamma \equiv \arg(-V_{ud}V_{ub}^*/V_{cd}V_{cb}^*)$ in the CKM unitarity triangle
        \item Explanation that tree-level $B$-meson decays provide theoretically clean measurements with negligible theoretical uncertainties
        \item Key decay channels: $B^\pm \to Dh^\pm$ (with various $D$ final states), $B^\pm \to D^*h^\pm$, $B^\pm \to DK^{*\pm}$, $B^0 \to DK^{*0}$, and $B_s^0 \to D_s^\mp K^\pm$ decays
        \item Basic methodology involving interference between $B$ decay amplitudes with different weak phases
        \item Recognition that $D$ meson final states must be accessible to both $D^0$ and $\overline{D}^0$ decays
    \end{itemize}
    \item \textbf{Expert-Level Requirements:}
    \begin{itemize}
        \item Discussion of different analysis techniques: GGSZ/ADS methods for multibody $D$ decays vs GLW method for CP eigenstates
        \item Treatment of hadronic parameter correlations and coherence factors in multibody decays
        \item Simultaneous fitting approach that combines beauty and charm sector measurements to constrain nuisance parameters
        \item Statistical methodology considerations including Feldman-Cousins confidence intervals and profile likelihood methods
        \item Cross-validation between different $B$ meson species ($B^\pm$, $B^0$, $B_s^0$) as BSM sensitivity test
        \item Importance of external constraints from CLEO-c and BESIII for $D$ decay strong-phase measurements
    \end{itemize}
    \item \textbf{Factual Benchmarks:}
    \begin{itemize}
        \item Current world-best $\gamma$ measurement precision is $\pm 2.8°$ from LHCb combination ($\gamma = 64.6 \pm 2.8°$)
        \item $B^\pm \to Dh^\pm$ with $D \to h^\pm h'^{\mp}$ provides the most statistically powerful $\gamma$ constraint
        \item Systematic uncertainties contribute $\sim 1.4°$ to total $\gamma$ uncertainty, indicating statistical dominance
        \item Key hadronic parameters: $r_{DK^\pm}^{B^\pm} \sim 9.7\%$, $\delta_{DK^\pm}^{B^\pm} \sim 127°$, with typical uncertainties of $0.2\%$ and $3°$ respectively
        \item $D^0$ mixing parameters $x = (0.41 \pm 0.05)\%$ and $y = (0.621 \pm 0.021)\%$ must be measured simultaneously
        \item Time-dependent vs time-integrated analyses show consistent $\gamma$ values within uncertainties
        \item Charm sector measurements contribute through $D^0 \to K^\pm \pi^\mp$, $D^0 \to K_S^0 \pi^+ \pi^-$, $D^0 \to h^+ h^-$ decay studies
        \item External inputs required include coherence factors $\kappa$, CP-even fractions $F_+$, and strong phase differences $\delta$ from quantum-correlated $D$ meson studies
        \item Statistical combination uses $\sim 200$ input observables to determine $\sim 50$ parameters with goodness-of-fit $\sim 20\%$
    \end{itemize}
\end{itemize}
\end{quote}

For this query the \GraphRAG finds the top 20 observables that are physical constants and ranks them by semantic similarity to "gamma, CKM angle gamma" using vector embeddings. For each of these gamma-related observables, it identifies the decay processes used to measure them, finds any uncertainty sources that affect the measurements (along with the average importance ranking of that uncertainty source), and collects information about the scientific papers that studied these observables. The query then aggregates this information to return a comprehensive view of each observable including its similarity score to gamma, the decay processes involved in measuring it, relevant scientific papers, and the top 5 most significant uncertainty sources ranked by their impact on the measurement precision, with results ordered by similarity to gamma from highest to lowest.

The following answers were constructed from the context provided by each of the three RAG systems:

\begin{quote}
\textbf{GraphRAG:}
Use a combination of tree-level $B \to D^{(*)} h$ modes plus the time-dependent $B_s \to D_s^{\mp}K^{\pm}$ channel. Practically this means including $B^{\pm} \to D K^{\pm}$ and $B^{\pm} \to D \pi^{\pm}$ (with $D \to K_S \pi \pi$, $K_S K K$, $K^{\pm} \pi^{\mp}$, $K^{\pm} \pi^{\mp} \pi^{\pm} \pi^{\mp}$, 3- and 4-body final states), $B^{\pm} \to D^* K^{\pm}$, $B^0 \to D K^{*0/\pm}$ and $B^0 \to D K^+ \pi^-$ (GLW/ADS/GGSZ-type approaches), and the time-dependent $B_s^0 \to D_s^{\mp} K^{\pm}$ analyses --- combining these channels is the route taken in recent LHCb determinations and gives the best precision on $\gamma$ [2010.08483, 2410.21115, 2209.03692, 2412.14074].

The dominant systematic uncertainties will be external hadronic inputs and model/theory assumptions: uncertainties on $D$-decay hadronic parameters ($r_D^X$, $\delta_D^X$, $\kappa_D^X$, CP-even fractions $F_+$ for multi-body modes) and the coherence/strong-phase inputs used in model-independent Dalitz/binning approaches (these appear as a separate ``external/strong-phase'' uncertainty in several measurements) will dominate; if you use U-spin relations the non-factorizable U-spin-breaking modelling is a large, non-reducible theory systematic; and using external constraints (e.g.\ world-average $\gamma$ or $-2\beta_s$ as inputs) propagates their uncertainties into your result [2010.08483, 1408.4368, 2311.10434].

\textbf{TreeRAG:}
Use $B \to D K$ family modes with both time-integrated and time-dependent approaches. In practice the most powerful channels in the provided context are $B^{\pm} \to D K^*(892)^{\pm}$ (with $D$ reconstructed in two-, three- and four-body final states: $K^{\pm}\pi^{\mp}$, $K^+K^-$, $\pi^+\pi^-$, $K^0_S \pi^+ \pi^-$, $K^0_S K^+ K^-$, $K^{\pm} \pi^{\mp} \pi^{\pm} \pi^{\mp}$, \ldots) and the standard $B^{\pm} \to D K^{\pm}$ channels using GGSZ/ADS/GLW variants; include $B^0 \to D K^{*0}$ and time-dependent $B_s^0 \to D_s^{(*)\mp} K^{\pm}$ analyses for complementary sensitivity to $\gamma$ [2410.21115, 2401.17934, 1605.01082, 1503.09086].

The dominant systematic uncertainties you should expect are: the $D$-decay amplitude model (or the limited strong-phase inputs when using model-independent binned GGSZ) which directly feeds into $\gamma$ (model uncertainties shown as a separate error in GGSZ analyses) [1605.01082]; resonance and $K^*$ modelling / hadronic parameters ($r_B$, $\delta_B$) in $K^*$ regions and associated model variations [2401.17934]; detector and analysis systematics relevant to time-dependent modes such as decay-time acceptance, time-resolution and flavour-tagging calibration, plus production/reconstruction asymmetries and efficiency/angular corrections [1903.05530]; and external normalisation/branching-fraction uncertainties when quoting absolute rates [1503.09086]. In the recent LHCb results these systematics are subleading to statistics but the $D$ amplitude / strong-phase and resonance-modelling terms are the ones to watch as you scale to larger data samples [2410.21115, 2401.17934].

\textbf{BaseRAG:}
Use $B \to D K^{(*)}$ tree-level modes that exploit interference between $b \to c$ and $b \to u$ amplitudes: in particular $B^{\pm} \to D K^{*\pm}$ (with $D$ decays to two-, three- and four-body final states) and $B^0 \to D K^{*0}$ (including $D \to K^0_S h^+ h^-$ to give strong-phase information) --- supplement with the benchmark $B^{\pm} \to D K^{\pm}$ channels and, where available, $B_s \to D \phi$ modes to add independent sensitivity [2410.21115, 2401.17934, 1709.05855, 1605.01082, 1308.4583]. Including the $D \to K^0_S h^+ h^-$ (model-independent) inputs breaks discrete degeneracies and improves precision [2401.17934, 1605.01082].

The dominant systematic sources seen in these analyses are external inputs and modelling/efficiency uncertainties: limited precision of normalization branching fractions (can be the largest single systematic) and limited simulation sample sizes that affect efficiency estimates [2108.07678, 2305.01463, 2011.00217]. Other important systematics are $D$-decay amplitude/strong-phase modelling (or the need for external binwise phase inputs), signal/background modelling and fit templates, detector/selection effects (track-reconstruction, trigger, reweighting), and flavour-tagging / decay-time acceptance and time-resolution uncertainties or resonance-modelling choices in amplitude fits [1903.05530, 2011.00217, 2305.01463, 1605.01082].
\end{quote}

The LLM evaluator rated the \GraphRAG response "Good", and the \TreeRAG and BaseRAG responses "Satisfactory" based on the evaluation rubric provided. These scores were validated against a human expert, which agreed with the ratings.

\section{\TreeRAG Implementation Details}
\label{app:TreeRAG}

\subsection{Paper Tree Schema}

The \TreeRAG system transforms each unstructured \LaTeX~source document in the corpus into a paper tree representation that preserves the logical organization of the document. Each node in the paper tree represents a semantic unit of the paper, with the following schema:

\begin{itemize}
\item \textbf{Node attributes:}
\begin{itemize}
\item \texttt{title}: Section or element identifier
\item \texttt{summary}: Summary text (paper abstract for root, LLM-generated summary for internal nodes, raw content for leaves)
\item \texttt{embedding}: The diffused dense vector representation combining the node's summary with its children summaries
\item \texttt{parent}: Reference to parent node (null for root)
\item \texttt{sections}: List of references to child nodes
\end{itemize}
\end{itemize}

\subsection{Paper Tree Construction}

The paper tree construction begins by identifying latex (sub)sections, which form branches in the tree. The text of a (sub)section above a defined length is automatically split into (sub)subsections, otherwise it is chunked and turned into leaf nodes. Atomic content units such as figures, tables, and equations are given their own leaf node, with their captions as abstracts to ensure that semantic search can locate specific visual or mathematical content within the paper.

The summarization process operates depth-first, beginning at the deepest level of each paper tree and moving upward. This ensures that parent summaries can incorporate the distilled content of their children into their own summaries. For leaf nodes, the summary is simply the node's text content (or caption for figures/tables). For internal nodes, the system concatenates all child summaries and generates a distilled summary.

This summarization can be performed asynchronously across all identical depth nodes within the corpus. For the LHCb corpus using GPT-5 nano 7049 summaries were generated in total with an average summary length of 249 tokens. This took 30 minutes of processing and cost 6 USD.

Dense vector embeddings are generated for each node summary with the 384 dimensional \texttt{BAAI/bge-small-en-v1.5} paragraph embedding model. The embedding diffusion process then refines these initial representations by incorporating relational information from within the tree structure. For each node in the tree, the algorithm constructs a diffused embedding as the sum of that node's original embedding and the attention weighted sum of its children's embeddings:

\begin{align}
\mathbf{e}'_v &= \lambda \mathbf{e}_v + (1-\lambda) \sum_{c \in \text{children}(v)} w_{v,c} \mathbf{e}_c \\
w_{v,c} &= \frac{\exp(\mathbf{e}_v \cdot \mathbf{e}_c \;/\;\tau))}{\sum_{c \in \text{children}(v)} \exp(\mathbf{e}_v \cdot \mathbf{e}_{c}) \;/\;\tau}
\end{align}

Here $\lambda$ is the diffusion parameter, which controls the extent to which the child embeddings should be diffused into the node embedding, and $\tau$ is the temperature parameter, which controls how biased the attention weighted sum should be towards child embeddings which are already semantically similar to the node embedding.

\subsection{\TreeRAG Context Retrieval}

\begin{algorithm}[H]
\caption{Context Retrieval Algorithm}
\label{alg:tree-search}
\begin{algorithmic}[1]
\REQUIRE Forest $\mathcal{F} = \{T_1, T_2, \ldots, T_n\}$ where each $T_i$ is a paper tree
\REQUIRE Target context size $k \in \mathbb{Z}^+$ (number of tokens)
\REQUIRE Query embedding $\mathbf{q} \in \mathbb{R}^d$
\STATE \textbf{Initialize:}
\STATE $\mathcal{B} \gets \{root(T_i) : T_i \in \mathcal{F}\}$ \COMMENT{Boundary set initialized with all roots}
\STATE $\mathcal{C} \gets \emptyset$ \COMMENT{The context is initially empty}
\STATE \textbf{Precompute:} $\forall \;\text{n} \in \mathcal{B}: s(\text{n}) \gets \text{similarity}(\text{n.embedding}, \mathbf{q})$

\WHILE{$|\mathcal{C}| < k$ \AND $\mathcal{B} \neq \emptyset$}
    \STATE $\text{n}^* \gets \arg\max_{\text{n} \in \mathcal{B}} s(\text{n})$ \COMMENT{Get most similar node in boundary set}
    \STATE $\mathcal{B} \gets \mathcal{B} \setminus \{\text{n}^*\}$ \COMMENT{Remove from boundary set}

    \IF{$\text{hasChildren}(\text{n}^*)$}
        \STATE $\mathcal{B} \gets \mathcal{B} \cup \text{n}^*\text{.children}$ \COMMENT{Add children to boundary}
        \STATE \textbf{Compute:} $\forall \text{c} \in \text{children}(\text{n}^*): s(\text{c}) \gets \text{similarity}(\text{c.embedding}, \mathbf{q})$
    \ELSE
        \STATE $\mathcal{C} \gets \mathcal{C} \cup \{\text{n}^*\}$ \COMMENT{Add leaf to relevant set}
    \ENDIF
\ENDWHILE
\RETURN $\mathcal{C}$
\end{algorithmic}
\end{algorithm}

The context for the \TreeRAG is constructed according to Algorithm \ref{alg:tree-search}. This algorithm implements a best-first search strategy across a forest of hierarchical document trees to retrieve the most relevant content for a given query. Starting with all tree roots in a boundary set, the algorithm iteratively selects the node with highest similarity to the query embedding, then either expands it by adding its children to the boundary (if it has children) or adds it to the final context set (if it's a leaf node). This greedy approach efficiently navigates multiple paper trees simultaneously, using semantic similarity to guide the search toward the most query-relevant leaf nodes.

The tree structure also allows for less fragmented context construction than a traditional RAG system. For example, the hierarchical organization means that leaf nodes from the same paper can be added to the context together, preventing the scattered mixing of content from different sources that often occurs in traditional RAG systems. Additionally, the tree structure enables the inclusion of relevant paper abstracts and higher-level contextual information if needed, providing the LLM with a coherent semantic context to surround each retrieved chunk. This ensures that related content can maintain logical relationships and that each piece of information is presented within its proper context, leading to more coherent and contextually-aware responses.

\section{\GraphRAG Implementation Details}
\label{app:GraphRAG}

\subsection{Knowledge Graph Schema}

The current knowledge graph implementation is meant to demonstrate the feasibility of automated structured knowledge extraction from physics literature via LLMs. Thus, it represents a prototype system rather than a comprehensive ontology of the domain. For this work the knowledge graph includes five primary entity types:

\begin{itemize}[leftmargin=1.5em]
    \item \textbf{paper}: Represents individual publications with attributes:
    \begin{itemize}
        \item Name (arXiv identifier)
        \item Description (abstract text)
        \item Data-taking period (run 1, run 2, run 3, ift)
        \item Analysis strategy (angular analysis, amplitude analysis, search, other)
        \item Embedding vector
    \end{itemize}

    \item \textbf{observable}: Physical quantities measured in analyses with attributes:
    \begin{itemize}
        \item Name (standard notation, e.g., $\mathcal{B}(B_s^0 \to \mu^+\mu^-)$)
        \item Description (natural language explanation)
        \item Type (branching fraction, branching ratio, physical constant, angular observable, functional dependence)
        \item Embedding vector
    \end{itemize}

    \item \textbf{decay}: Particle decay processes with attributes:
    \begin{itemize}
        \item Name (PDG notation, e.g., "$B_s^0 \to D_s^{\mp}K^{\pm}$")
        \item Parent particle
        \item Children particles (decay products)
        \item Production mechanism (p-p, Pb-Pb, p-Pb, Xe-Xe, O-O, Pb-Ar, p-O)
        \item Embedding vector
    \end{itemize}

    \item \textbf{uncertainty\_source}: Sources of systematic uncertainty with attributes:
    \begin{itemize}
        \item Name (standardized identifier)
        \item Description (detailed explanation)
        \item Type classification (statistical, internal systematic, external systematic)
        \item Embedding vector
    \end{itemize}

    \item \textbf{method}: Analysis techniques for uncertainty estimation with attributes:
    \begin{itemize}
        \item Name (technique identifier)
        \item Description (implementation details)
        \item Embedding vector
    \end{itemize}
\end{itemize}

Entities in the graph are connected through four typed relationships:

\begin{itemize}[leftmargin=1.5em]
    \item \textbf{determines}: paper → observable (the paper determines this observable)
    \begin{itemize}
        \item Value (the measured value of the observable with uncertainties)
    \end{itemize}
    \item \textbf{measured\_with}: observable → decay (the observable is measured using this decay channel)
    \item \textbf{affects}: uncertainty\_source → observable (this source of uncertainty affects the measurement of an observable) with attributes:
    \begin{itemize}
        \item Ranking (integer importance score, 1 = most significant)
        \item Magnitude (numerical contribution when quantified)
        \item Condition (applicability constraints or context)
    \end{itemize}
    \item \textbf{estimates}: method → uncertainty\_source (this method is used to evaluate the uncertainty)
\end{itemize}

\begin{figure}[H]
    \centering
    \includegraphics[width=0.45\linewidth, trim=0 10pt 0 0, clip]{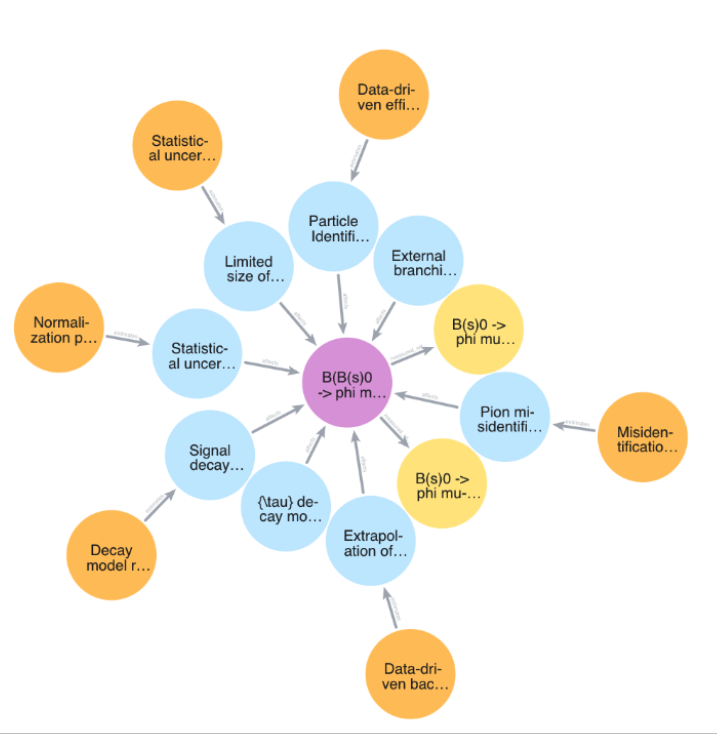}
    \includegraphics[width=0.45\linewidth]{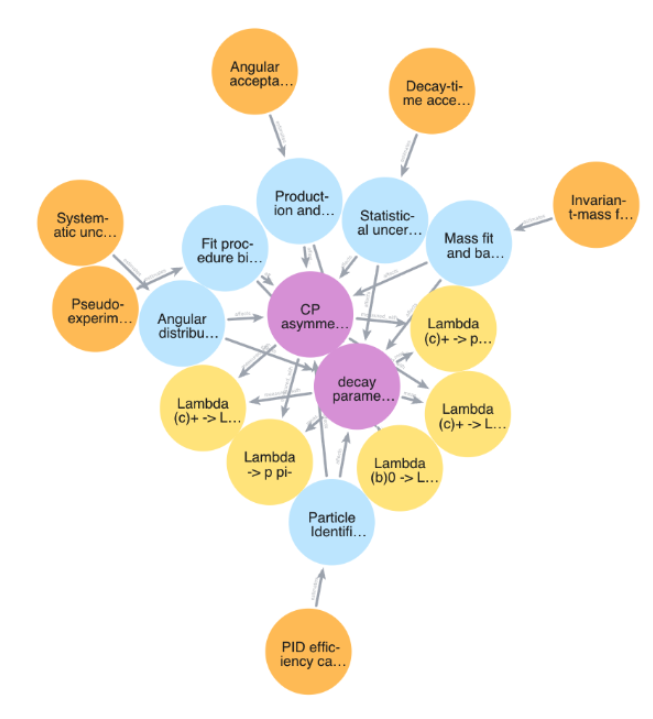}
    \caption{Example per-paper knowledge graphs showing entity types and relationships. Purple entities are observables, yellow entities are decays, light blue entities are sources of uncertainty, orange entities are analysis methods. Left: Analysis of $B_s^0 \to D_s^{\mp}K^{\pm}$ decay time-dependent CP violation. Right: Study of $\Lambda_b^0$ production asymmetry.}
    \label{fig:kg-examples}
\end{figure}

\subsection{Knowledge Graph Construction}

\subsubsection{Entity Extraction}

Entity extraction employs a two-stage LLM-based approach that processes different sections of each paper to gather complementary information about the analysis methodology and systematic uncertainties.

\paragraph{Abstract Processing} The abstract provides high-level metadata about the analysis strategy and physical observables. A prompt with few-shot examples extracts:
\begin{itemize}
    \item Primary observable(s) with type classification (branching fractions, angular observables, physical constants, etc.)
    \item Decay channels with standardized particle names
    \item Data-taking period (Run 1, Run 2, Run 3, or heavy-ion collisions)
    \item Analysis strategy classification (angular analysis, amplitude analysis, search, or other precision measurements)
\end{itemize}

The abstract processing uses domain-specific normalization rules to ensure consistency. For example, ``$B_s \to \mu^+ \mu^-$'' is automatically mapped to ``B(s)0 -> mu+ mu-'', while production mechanisms are inferred from collision energy and particle types mentioned in the text.

\paragraph{Systematic Uncertainties Processing} To handle the computational constraints of processing full papers, the system employs targeted text pre-processing:
\begin{itemize}
    \item Automated identification of systematic uncertainty sections using keyword matching (``error'', ``uncertain'', ``systematic'')
    \item Removal of non-analytical content (introduction, detector descriptions, acknowledgments)
    \item Prioritization of content near uncertainty tables and error budget discussions
\end{itemize}

After isolating relevant sections, an LLM prompt extracts the remaining entity types from the paper body. The previously extracted observables from the abstract are provided in the context:
\begin{itemize}
    \item \textbf{Uncertainty sources}: Each source receives a three-way classification (statistical, internal systematic, or external systematic) based on whether the collaboration can directly control or improve the uncertainty through analysis choices
    \item \textbf{Methods}: Specific techniques used to quantify systematic effects, interested in transferable methodologies applicable across different analyses
\end{itemize}

The extraction process simultaneously identifies relationships between uncertainty sources and the pre-extracted observables, capturing exact magnitude values as reported in the paper.

The prompt design incorporates several strategies validated during development. The LLM generates explanations before structured output to ensure comprehensive understanding of extraction requirements. Additionally, entities and their relationships are extracted simultaneously rather than in separate passes, reducing context switching and minimizing potential misalignment between uncertainty sources and their quantitative impacts on specific observables.

After processing both abstract and full-text content, the system constructs complete paper entities using the metadata extracted from the abstract (data-taking period, analysis strategy, etc.) and creates relationships linking papers to their determined observables. All extracted information is validated against the knowledge graph schema and stored in a Neo4j graph database with full provenance tracking through arXiv identifiers.

For the LHCb corpus of 837 papers, the full graph extraction process took 2.5 hours and cost 18 USD with GPT-5 mini.

\subsubsection{Entity Canonicalization}

The per-paper extraction process produces a corpus-wide knowledge graph that requires cross-document entity resolution within each entity class to identify when different papers reference identical concepts despite variations in terminology, notation, or descriptive language.

\paragraph{Similarity-Based Clustering} The canonicalization pipeline begins by creating hybrid representations for each entity, combining TF-IDF vectors of their names with semantic embeddings of their descriptions. Entity names are vectorized using TF-IDF with character n-grams to capture lexical variations, while descriptions are encoded using the 384-dimensional \texttt{BAAI/bge-small-en-v1.5} paragraph embedding model to capture semantic content. These complementary vector representations are turned into cosine similarity matrices and their weighted sum is taken to produce the final similarity scores for clustering.

The resulting similarity matrix drives an agglomerative clustering algorithm with dynamic threshold adjustment. The system iteratively reduces similarity thresholds until cluster sizes remain below a configurable maximum, ensuring that cluster sizes remain manageable for downstream LLM processing.

\paragraph{LLM-Guided Merge Decisions} Each cluster containing multiple entities undergoes evaluation by an LLM judge that determines which entities represent identical concepts and should truly be merged. The LLM can choose to split algorithmically-generated clusters when it determines that entities are related but conceptually distinct---for example, distinguishing ``tracking efficiency in the vertex detector'' from ``tracking efficiency in the muon system'' despite their semantic similarity. For entities deemed identical, the LLM synthesizes a unified name and description that captures the essential information from all merged entities.

This hybrid clustering + LLM based approach combines similarity clustering for computational efficiency with the expanded reasoning capabilities of an LLM, in order to achieve scalable processing on large graphs while maintaining nuanced decisions requiring domain knowledge.

\paragraph{Iterative Refinement} The canonicalization process operates iteratively, with each round potentially enabling new merging opportunities as the entire graph evolves. The process continues until a configurable stopping criterion is met---typically when fewer than 10\% of entities are merged in a given iteration, indicating that the major canonicalization opportunities have already been identified and resolved.

For the LHCb corpus of 837 papers, this entire process took 45 minutes and cost 8 USD with GPT-5 nano. The canonicalization process achieved significant entity deduplication: uncertainty\_source entities were reduced from 7167 to 2895 (60\% reduction), method entities from 6792 to 1786 (74\% reduction), observable entities from 2166 to 2028 (6.4\% reduction), and decay entities from 1595 to 1495 (6.3\% reduction). Paper entities were not considered for canonicalization as they represent unique documents rather than potentially duplicate concepts.

\subsection{Graph RAG Context Retrieval}

The Graph RAG query interface enables natural language querying of the knowledge graph through a multi-stage pipeline that translates questions into \textsc{Cypher} queries, executes them against the Neo4j database, and synthesizes natural language responses from the structured results.

\paragraph{Natural Language to \textsc{Cypher} Translation} Natural language queries are translated into \textsc{Cypher} graph database queries using an LLM equipped with comprehensive knowledge of the graph schema and particle physics domain terminology. The translation prompt incorporates detailed schema documentation including all node types (paper, observable, decay, uncertainty\_source, method) with their properties, relationship types with attributes, and standardized particle naming conventions from the Particle Data Group.

The system uses few-shot prompting with three representative examples covering common query patterns: method identification for specific systematic uncertainties, frequency analysis of uncertainty sources across decay types, and comprehensive uncertainty assessment for physics measurements. Each example demonstrates proper \textsc{Cypher} syntax, semantic similarity usage, and result aggregation techniques.

\paragraph{Query Construction Principles} The LLM is instructed to follow these principles when generating \textsc{Cypher} queries:

\begin{itemize}
\item \textbf{Semantic Similarity Search}: Rather than exact string matching, the system uses embedding-based similarity for flexible entity retrieval. Descriptive queries are converted to embeddings using the same sentence transformer model employed during knowledge graph construction.

\item \textbf{Result Aggregation}: To provide meaningful answers while avoiding redundancy, queries aggregate related information using \textsc{Cypher} aggregation operations. The results are limited to prevent context overflow in downstream synthesis.

\item \textbf{Ranking and Ordering}: Queries incorporate domain-appropriate ranking metrics such as uncertainty importance rankings, frequency of occurrence across analyses, or semantic similarity scores to prioritize the most relevant results.

\item \textbf{Provenance Preservation}: Every query returns relevant \texttt{arxiv\_id} values from associated papers to enable source citation and verification of results.

\item \textbf{Query Justification}: Before generating the \textsc{Cypher} query, the LLM must provide an "explanation" field that describes the query logic, traversal path, and ranking strategy. This forces the model to articulate its reasoning process, leading to more thoughtful and accurate query construction while providing transparency for debugging and validation.
\end{itemize}

\paragraph{Query Execution and Embedding Integration} The generated \textsc{Cypher} queries are preprocessed to handle semantic similarity operations and the processed \textsc{Cypher} queries are executed against the Neo4j database using the py2neo Python driver, which returns structured results as cursor objects that are converted to dictionary format for downstream processing.

\end{document}